\documentclass[aip,apl,amsmath,amssymb,amsfonts,reprint,numerical]{revtex4-1}

\usepackage{graphicx}
\usepackage{dcolumn}
\usepackage{bm}
\usepackage{epsf}
\usepackage[usenames]{color}

\usepackage{amsmath}

\newcommand{\be}{\begin{equation}}
\newcommand{\ee}{\end{equation}}
\newcommand{\bea}{\begin{eqnarray}}
\newcommand{\eea}{\end{eqnarray}}
\newcommand{\Eq}[1]{Eq.\,(\ref{#1})}% \Eq{abc}
\newcommand{\Fig}[1]{Fig.\,\ref{#1}}% \Fig{fig:abc}

\begin{document}

\title{Terahertz emitters based on microcavity dipolaritons}

\author{A. Seedhouse}
\affiliation{School of Physics and Astronomy, Cardiff University, Cardiff CF24 3AA, United Kingdom}
\author{J. Wilkes}
\affiliation{School of Physics and Astronomy, Cardiff University, Cardiff CF24 3AA, United Kingdom}
\author{V. D. Kulakovskii}
\affiliation{Institute of Solid State Physics, Chernogolovka 142432, Russia}
\affiliation{National Research University Higher School of Economics, \mbox{20\,Myasnitskaya street,\,Moscow\,101000,\,Russia}}
\author{E. A. Muljarov}
\affiliation{School of Physics and Astronomy, Cardiff University, Cardiff CF24 3AA, United Kingdom}

%\date{\today}

\begin{abstract}
We propose the use of dipolaritons -- quantum well excitons with large dipole moment, coupled to a planar microcavity -- for generating terahertz (THz) radiation. This is achieved by exciting the system with two THz detuned lasers that leads to dipole moment oscillations of the exciton polariton at the detuning frequency, thus generating a THz emission. We have optimized the structural parameters of a system with microcavity embedded AlGaAs double quantum wells and shown that the THz emission intensity is maximized if the laser frequencies both match different dipolariton states. The influence of the electronic tunnel coupling between the wells on the frequency and intensity of the THz radiation is also investigated, demonstrating a trade-off between the polariton dipole moment and the Rabi splitting.
\end{abstract}

\maketitle

Terahertz (THz) waves can propagate through materials without ionizing atoms, which leads to a variety of applications, such as security screening and medical imaging.\cite{Dean2014} However, due to the reduced interaction of the THz radiation with matter, generating THz waves is also harder to achieve
%compared to radiation at shorter wavelengths,
that makes it presently one of the grand technological challenges.\cite{SiegelIEEE02} To tackle this challenge, bosonic cascade lasers based on excitonic transitions in parabolic traps inside microcavities (MCs) have been proposed as sources of THz radiation.\cite{LiewPRL13} Furthermore, electromagnetically induced transparency in the THz range of such bosonic cascades  was also demonstrated.\cite{LiewOL18}
More `traditional' sources of THz radiation include Gunn devices and tunnel diodes,\cite{EiseleIEEE00} quantum cascade lasers,\cite{FaistSci94,KohlerNat02} free-electron based sources,\cite{GoldRSI97} as well as laser-driven emitters of Cherenkov radiation in the THz range.\cite{AustonPRL84}

\begin{figure}[t]
%    \centering
\vskip4mm
\hskip-2mm
\includegraphics*[width=0.15\textwidth,angle=0]{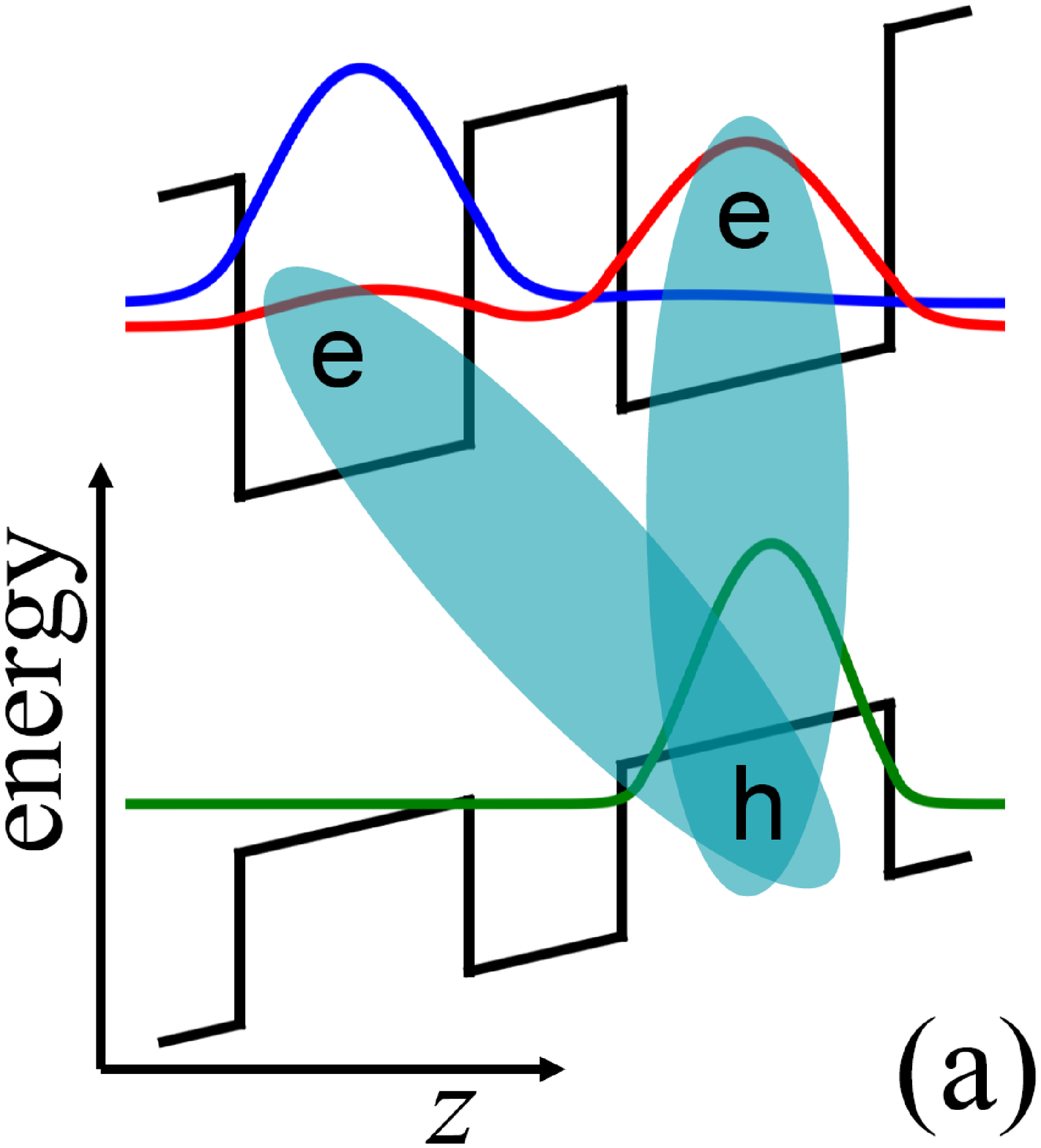}
%\vskip1mm
\hskip5mm
\includegraphics*[width=0.30\textwidth,angle=0]{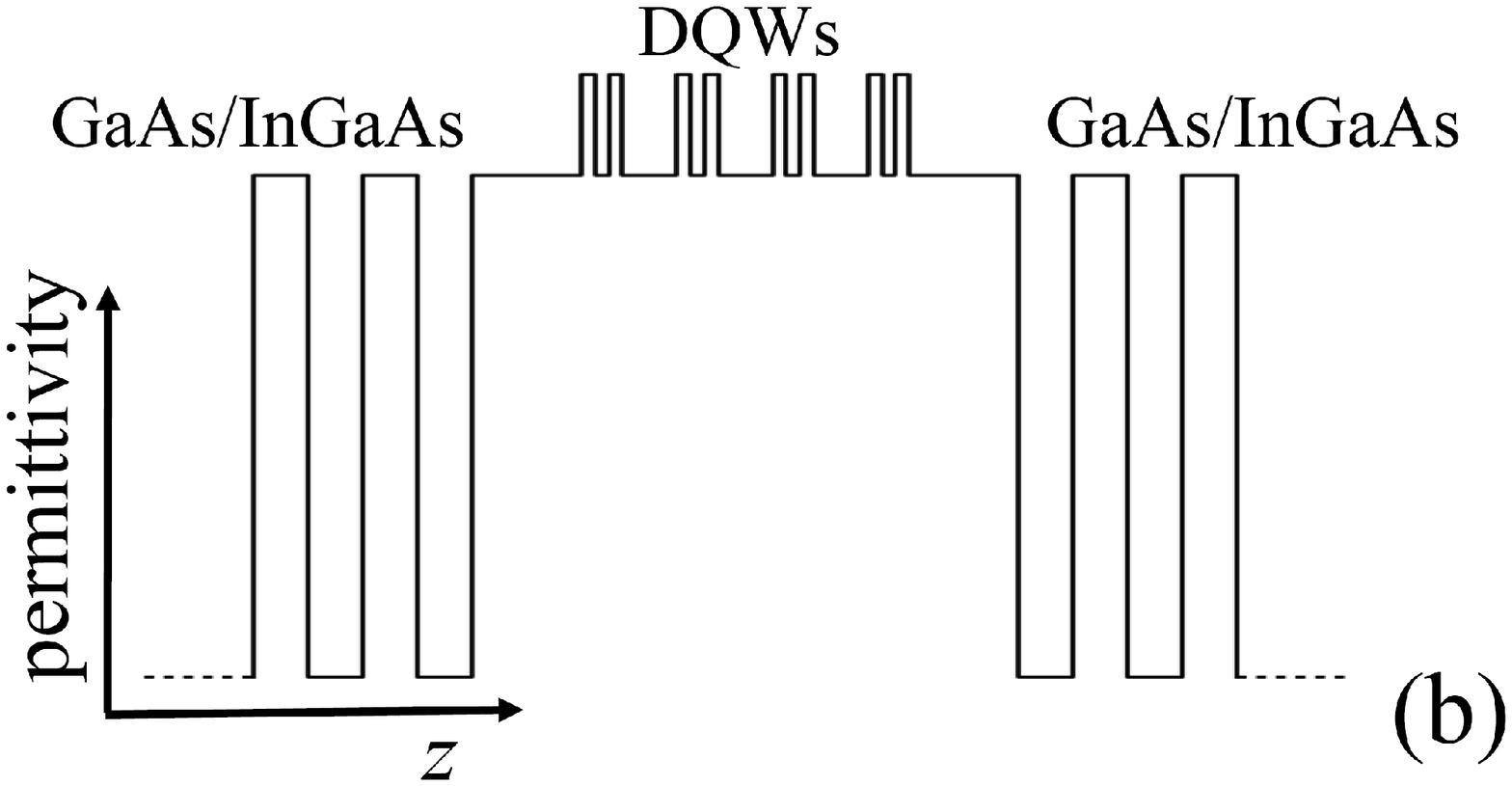}
\vskip-1mm
    \caption{(a) Wave functions of the electron ground (red) and excited state (blue), hole ground state (green), and their potentials (black) in a DQW with a static electric field applied in the growth direction. The direct (indirect) exciton is shown by the blue shading over the electron and hole in the same (different) QWs. (b) Permittivity as function of the growth coordinate $z$ of the MC with four DQWs placed in the antinode positions of the cavity mode.}
    \label{fig:schematics}
\end{figure}

Recently, it has been shown that systems containing exciton polaritons can provide an efficient tunable source of THz radiation,\cite{KyriienkoPRL13,KristinssonPRB13} and a THz laser using dipolaritons has been proposed.\cite{KristinssonPRA14} This proposal is based on a recent experimental observation of MC polaritons with a large dipole moment, called dipolaritons.\cite{cristofolini2012coupling} Qualitatively, a dipolariton presents a mixture of a cavity photon and tunnel-coupled bright direct and dark indirect excitons.\cite{cristofolini2012coupling,SivalertpornPRL15} Such exciton states  exist in a semiconductor double quantum well (DQW) system, as illustrated by Fig.\,\ref{fig:schematics}(a) showing that an electron and a hole localized in adjacent quantum wells (QWs) form an indirect exciton, while the oppositely charged carriers within the same QW are bound into a direct exciton. Owing to the charge separation, indirect excitons are characterized by an appreciable dipole moment $\sim d_0$ ($d_0$ is the center-to-center distance between the QWs) and a dramatically reduced oscillator strength and increased lifetime, as compared to the direct exciton.\cite{AlexandrouPRB90,SivalertpornPRB12}
In such systems, the indirect exciton component of the dipolariton can form an oscillating macroscopic dipole moment, capable of producing a secondary emission in the THz range.

Such a coherent secondary emission can be generated by a resonant excitation of the cavity mode (CM) that induces indirect exciton density oscillations at the frequency of the polariton Rabi splitting lying in the THz range.\cite{KyriienkoPRL13} The frequency of the dipole oscillations can be efficiently tuned by an applied voltage changing the polariton level structure,\cite{KyriienkoPRL13,KristinssonPRB13} and the field of the THz radiation can be further modulated and optimized by an applied ac-voltage.\cite{LiEPL14} In practice, however, this scheme would suffer from a significant disadvantage: The excitation field (resonant to the CM) would inevitably excite the whole spectrum of polariton states leading to multiple-frequency beatings in the generated THz signal. Indeed, even an approximate three-level model describing dipolariton, which was introduced in Ref.\,\onlinecite{cristofolini2012coupling}, shows three polariton branches always present in the spectrum, while an accurate microscopic calculation going beyond this simple picture demonstrates coexistence of a large number of dipolariton states.\cite{SivalertpornPRL15,WilkesPRB16}

In this Letter, we propose a reliable scheme for generating THz emission of dipolaritons by exciting MC embedded DQWs with two coherent lasers having a THz frequency detuning $\Delta$. We demonstrate theoretically that such a THz radiation is maximized and can reach reasonable intensities if the lasers are resonant to different polariton states. We focus in this work on a large class of AlGaAs DQW structures embedded in planar Bragg-mirror MCs, similar to those studied experimentally in Refs.\,\onlinecite{ChristmannAPL11} and \onlinecite{cristofolini2012coupling} (see Fig.\,\ref{fig:schematics}(b)), and provide their optimization with respect to structural and tuning parameters.
%We show in particular that ... {\bf Can add some more later}.

Our approach is based on the microscopic theory of polaritons in MC embedded multiple QWs in external electric and magnetic fields.\cite{SivalertpornPRL15,WilkesPRB16,WilkesNJP16,WilkesSM17} We solve Maxwell's wave equation for the optical field simulteneously with inhomogeneous Schr\"odinger's equation describing  optically driven excitons in DQWs in the presence of static external fields. The Schr\"odinger problem is treated by using a multi-sublevel approach\cite{SivalertpornPRB12,WilkesNJP16} which takes into account the role of different electron-hole pair states in the exciton formation, in this way providing an accurate calculation of a large number of excitonic states and properly describing their contribution to the optical response of the DQW system.

To find the intensity of the THz radiation emerging from a continuous wave excitation of the system by two laser beams with frequencies $\omega_1$ and $\omega_2$, we introduce a microscopic linear excitonic polarization, or the polariton wave function,
\begin{equation}
    Y(\boldsymbol{ \rho }; z_{e}, z_{h}; t) = Y_{1}e^{-i\omega_{1}t} + Y_{2}e^{-i\omega_{2}t}\,,
    \label{eq:dipole}
\end{equation}
in which $Y_j(\boldsymbol{ \rho }; z_{e}, z_{h})$ is the linear polarization due to a single-laser excitation at frequency $\omega_j$.
This polarization can be obtained as\cite{SivalertpornPRL15,WilkesPRB16}
\begin{equation}
    Y_j(\boldsymbol{ \rho }, z_{e}, z_{h}) = \int \chi(\boldsymbol{ \rho },z_{e}, z_{h};z; \omega_j) {\cal E}(z;\omega_j) dz\,,
\end{equation}
where ${\cal E}(z; \omega)$ is the macroscopic electric field within the optical system at frequency $\omega$ and $\chi(\boldsymbol{ \rho },z_{e}, z_{h}; z; \omega)$ is the microscopic non-local excitonic susceptibility. The
latter can be found from the spectral representation of the excitonic Green's function,
\begin{equation}
    \chi(\boldsymbol{ \rho },z_{e}, z_{h}; z; \omega) = d_{cv} \sum_{\nu} \frac{\Psi_{\nu}(\boldsymbol{ \rho }, z_{e}, z_{h}) \Psi_{\nu}(0, z, z)}{E_{\nu} - \hbar \omega - i\gamma}\,,
\end{equation}
where  $d_{cv}$ is the conduction-to-valence band dipole moment (we use $d_{cv} = 0.6$\,nm for GaAs), $\gamma$ is a phenomenological damping (we use $\gamma=0.1\,$meV), $E_{\nu}$ and $\Psi_{\nu}(\boldsymbol{ \rho }, z_e, z_h)$ are, respectively, the eigenenergy and the wave function of excitonic state $\nu$ in the DQW system without laser excitation, $\boldsymbol{ \rho}_{e(h)}$ and $z_{e(h)}$ are, respectively, the in-plane and perpendicular electron (hole) coordinates, and $\boldsymbol{ \rho }=\boldsymbol{ \rho }_e-\boldsymbol{ \rho }_h$.
The electric field ${\cal E}(z;\omega)$ is in turn found by solving Maxwell's wave equation with the actual permittivity profile of the layered structure, as shown in Fig.\,\ref{fig:schematics}(b), and a regularized nonlocal macroscopic susceptibility:\cite{WilkesPRB16} $\chi_{\omega}(z,z')=\chi(0,z, z;z';\omega) - \chi(0,z, z; z';0)$. This calculation thus takes into account the changes of the dielectric constant due to the alternating layers within the MC and due to the DQWs with the resonant excitonic contribution to the permittivity.

The presence of an indirect exciton component in the wave function $Y$ results in a non-vanishing macroscopic polariton dipole moment which can be found as\cite{SivalertpornPRL15}
\begin{equation}
    \mathcal{D}(t) \approx \mathcal{D}_{1} + \mathcal{D}_{2} + 2d_{12} \cos ( \Delta t + \phi)\,,
    \label{eq:dipoleMom2freq}
\end{equation}
where $\mathcal{D}_{1}$ and $\mathcal{D}_{2}$ are the static components of the dipole moment due to the individual polarization fields $Y_1$ and $Y_2$, respectively, $\phi$ is the phase difference between $Y_{1}$ and $Y_{2}$, and the last term in \Eq{eq:dipoleMom2freq}, oscillating with the detuning frequency $\Delta=\omega_2-\omega_1$, is the result of interference of the two polarizations, with
\be
    d_{12} = \frac{1}{\mathcal{N}} \!\int \!|Y_{1}(\boldsymbol{ \rho }, z_{e}, z_{h}) Y_{2}^{*}(\boldsymbol{ \rho }, z_{e}, z_{h})|
    (z_{e} -z_{h}) d \boldsymbol{ \rho } dz_{e} dz_{h}
\label{eq:dipolePol}
\ee
determining a normalized amplitude of the oscillating dipole moment. It is convenient to normalize the dipole moment in such a way that $\mathcal{D}_{j}$ provides the dipole moment of an individual polariton excited at frequency $\omega_j$. Therefore, $\mathcal{N}\approx \mathcal{N}_1+\mathcal{N}_2$ where $\mathcal{N}_j=\int |Y_j|^2d \boldsymbol{ \rho } dz_{e} dz_{h}$ is the number of polaritons  excited at frequency  $\omega_j$. The time-averaged radiation power $P$ due to the oscillating part of the polariton dipole moment can then be found from the formula for a classical oscillating dipole,\cite{LL2}
\begin{equation}
P = 4\mathcal{N}_{1}\mathcal{N}_{2}\frac{e^{2} d_{12}^{2} \Delta^{4}}{3  c^{3}}=P_0 \left(\frac{d_{12}}{d_0}\right)^2\left(\frac{\Delta}{\Delta_0}\right)^4,
    \label{eq:singleXpower}
\end{equation}
where $e$ is the electronic charge and $c$ is the speed of light.
Clearly, the radiation power is proportional to the product $\mathcal{N}_{1}\mathcal{N}_{2}$ of the numbers of polaritons excited by each laser, reflecting the coherent superposition of the two polarization waves.  Using a realistic estimate\cite{ButovNat02} for the maximum concentration $n_X=5\times 10^{10}$\,cm$^{-2}$ of indirect excitons excited in a DQW, we find the total number of polaritons (with the indirect exciton fraction of about 1/2) to be ${\cal N}\approx4\times 10^7$, within a laser excitation spot of about $0.1\,$mm in diameter, in a sample containing 4 DQWs. Assuming ${\cal N}_1={\cal N}_2={\cal N}/2$, we then obtain from \Eq{eq:singleXpower} an estimate for the THz radiation power $P_0=0.8\,$mW, for dipole moment $d_0=10.5$\,nm and frequency $\hbar\Delta_0=4.1$\,meV (equivalent to 1\,THz).

The modeled system is made up of four GaAs/Al$_{0.33}$Ga$_{0.67}$As DQWs placed at the antinode positions within a 5$\lambda$/2 MC, as it was done e.g. in Ref.\,\onlinecite{ChristmannAPL11}. The cavity consists of 17 and 21 pairs of alternating GaAs and InGaAs $\lambda$/4 layers forming distributed Bragg reflectors, see \Fig{fig:schematics}(b). The number of pairs determine the quality factor (Q-factor) of the cavity mode $Q\approx 7000$, which we have taken as a nearly fixed parameter. There are a large number of DQW systems that we explored, which are characterized by the widths of the barrier and wells. We found that the optimized structure for the chosen materials has the dimensions (well-barrier-well) 6-3-7\,nm, which we call System 2 (S2). A DQW with a wider barrier, 6-4-7\,nm, called S3, is also used for a detailed comparison. Furthermore, we demonstrate below data for the maximum emission power for three other DQW structures: 5-2-7\,nm (S1), 7-4-7\,nm (S4), and 7-4-8\,nm (S5).

We have studied the properties of excitons and polaritons, and the THz emission from such systems as functions of the cavity mode position $E_C$, the laser detuning frequency $\Delta$, and the applied external electric field $F$. In our calculation we have also added a small magnetic field of 1T, in order achieve a natural discretization the excitonic continuum, owing to formation of Landau levels,\cite{ButovPRL01,WilkesPRB16,WilkesNJP16} without having any sensible effect on the properties of the lower polariton states.

\begin{figure}[t]
\hskip-4mm
\includegraphics[width=0.50 \textwidth,angle=-90]{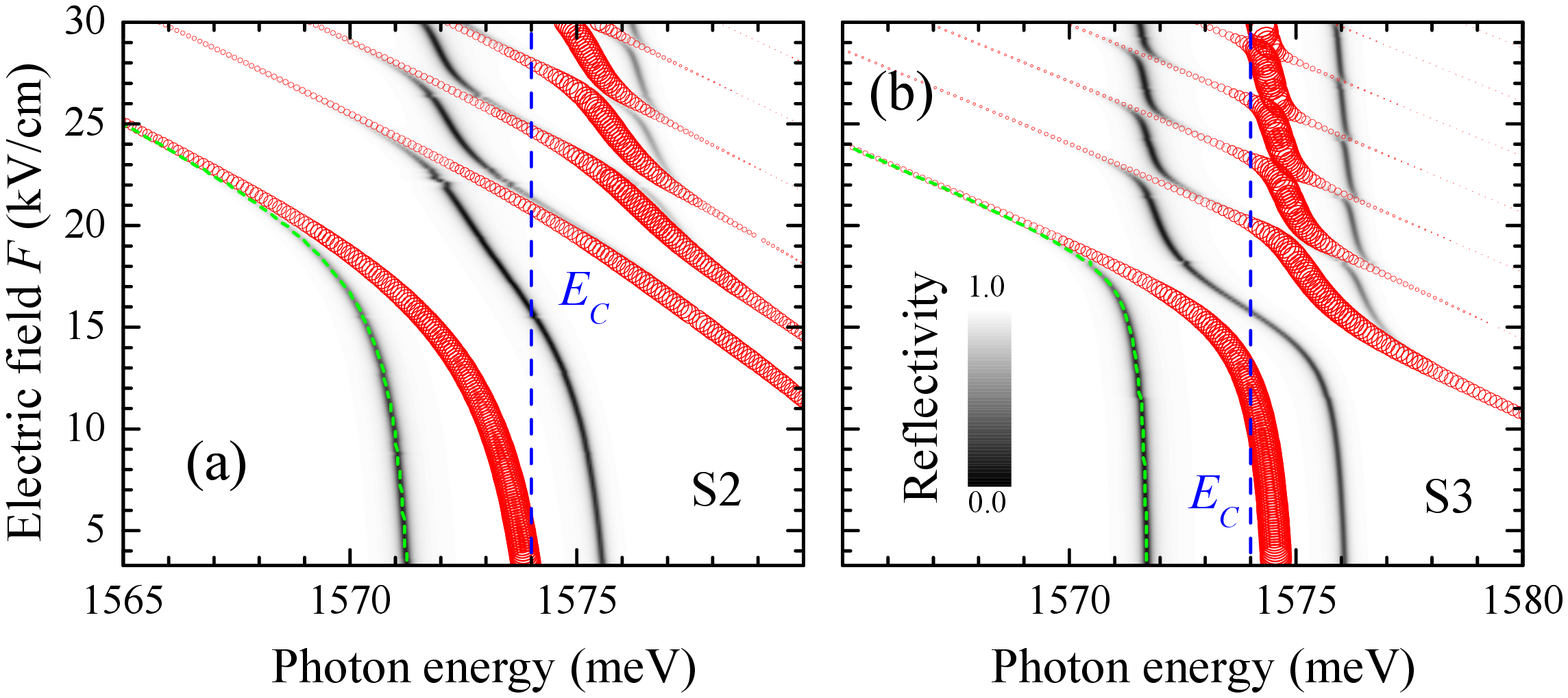}
\vskip-50mm
    \caption{Reflectivity spectra (gray scale map) of the DQW structure 6-3-7\,nm (S2, a) and 6-4-7\,nm (S3, b) for the bare cavity mode at $E_C=1574$\,meV (blue dashed line). The red circles show the exciton energies with the circle area being proportional to the exciton oscillator strength. The green short dashed lines follow the polariton ground state.}
    \label{fig:refl}
\end{figure}

Figure \ref{fig:refl} shows the exciton energies (red circles) and oscillator strengths (proportional to the circle area) in S2 and S3 DQWs, as well as the polariton reflectivity spectra in these systems (grey-scale) for the cavity mode at $E_C=1574$\,meV and varying electric field $F$. The dependence on $F$ of the exciton energy clearly demonstrates the relative contribution of the direct and indirect components in the exciton, which are causing, respectively, slow and quick (almost linear) changes of the exciton energy with $F$, owing to quite different separation of an electron and a hole within the DQW. The oscillator strength, providing exciton coupling to the optical cavity mode, in turn, demonstrates a transformation between the direct and indirect exciton, the latter being characterized by a dramatically reduced values due to a reduced overlap of the electron and hole wave functions. With decreased tunnel probability for a larger barrier in S3, such a crossover from direct to indirect exciton is becoming more prominent, manifesting itself in a sharper anticrossing well seen in \Fig{fig:refl}(b) near $F=17$\,kV/cm. Owing to their direct component, excitons are strongly coupled to the cavity mode, forming Rabi-split exciton-polariton states which are clearly seen in \Fig{fig:refl} as minima of the reflectivity. The presented results demonstrate multiple exciton and polariton states in the system which goes well beyond the simple three-level model of dipolaritons\cite{cristofolini2012coupling} (consisting of only direct, indirect exciton, and the cavity mode).

\begin{figure}[t]
%    \centering
\vskip3mm
\hskip-30mm
\includegraphics[width=0.5\textwidth,angle=-90]{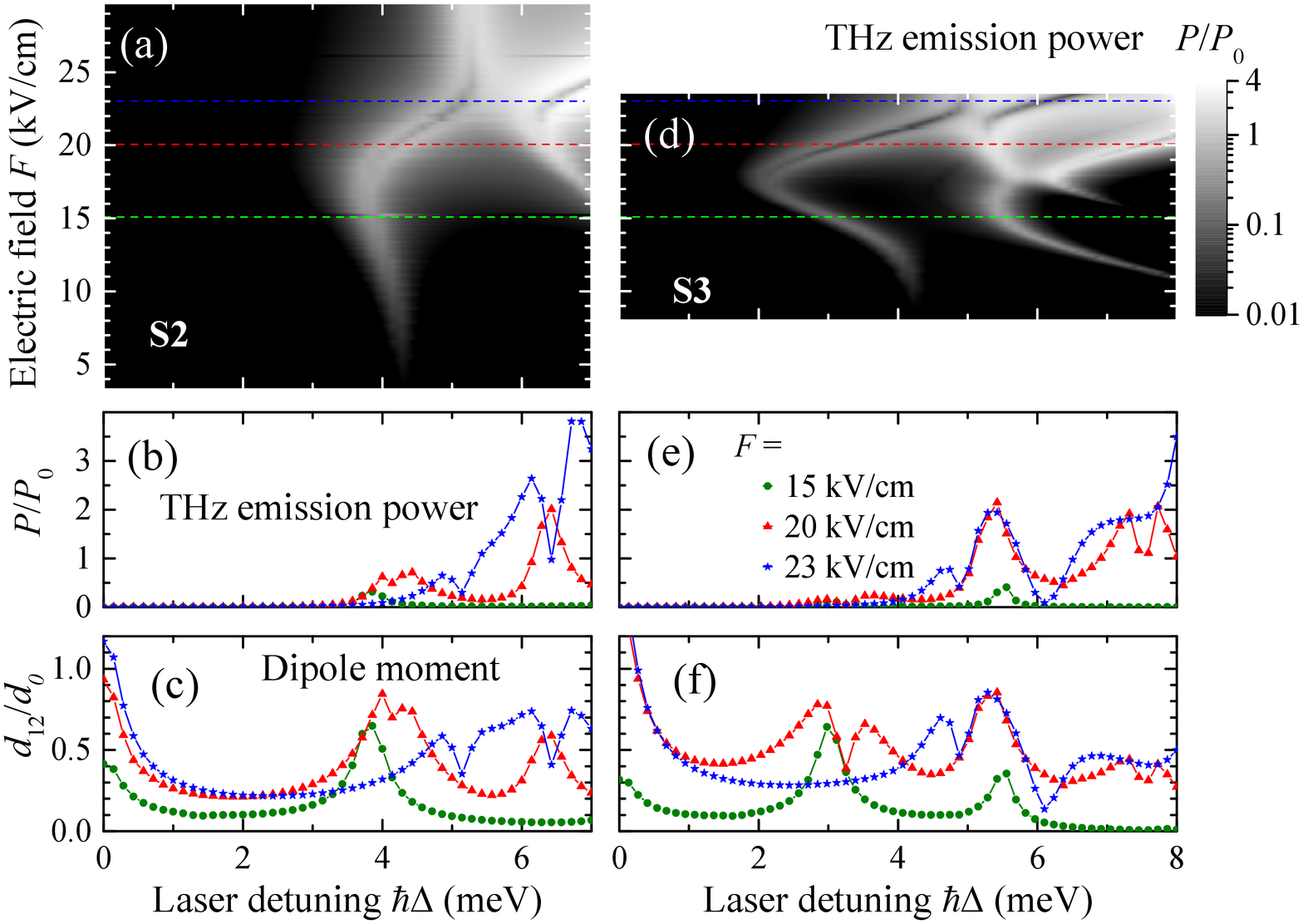}
\vskip-28mm
    \caption{Relative intensity ${P/P_0}$ (log scale) of the emission from the DQW system S2 (a) and S3 (d) as function of the emission frequency $\Delta$ and the electric field $F$, for the cavity mode at $E_C=1574$\,meV. The first laser frequency is matching the polariton ground state. (b,e) Relative intensity $P/P_0$ (linear scale) and (c,f) dipole moment $d_{12}/d_0$ as functions of the emission frequency $\Delta$ for $F=15.1$, 20.1, and 23.0\,kV/cm.}
    \label{fig:int}
\end{figure}
%    \caption{As Fig.\,\ref{fig:int1} but for the DQW system S3 (6-4-7\,nm).}

Figures \ref{fig:int}(a) and (d) show the relative power $P/P_0$ of the THz emission as a function of $F$ and the detuning $\Delta=\omega_2-\omega_1$ between laser frequencies, for systems S2 and S3, respectively, where $P_0$ is the power at the detuning of 1\,THz and $d_{12}=d_0=10.5$\,nm. The lower laser frequency $\omega_1$ is chosen to follow the lowest polariton state (green short dashed lines in \Fig{fig:refl}), otherwise the radiation intensity is significantly reduced. While the dipole moment $d_{12}$ shown in Figs.\,\ref{fig:int}(c) and (f) is maximized at around the energy of the ground polariton state ($\Delta=0$), the emission drops dramatically as $\Delta\to0$ due to the $\Delta^4$ factor in \Eq{eq:singleXpower}. The trade-off
between $d_{1,2}$ and $\Delta$ results in the maximum of $P$ to occur when the second laser frequency $\omega_2$ matches the frequency of an excited polariton state, for which $d_{12}$ shows a maximum, see \Fig{fig:int}(c). While higher polariton states produce lower $d_{12}$, the emission power due to such states can be higher, again because of the $\Delta^4$ factor, as it is clear from Figs.\,\ref{fig:int}(b) and (e) demonstrating the laser detuning dependencies for different values of the electric field.

Comparing THz emission from the two structures, S2 and S3, we see that the smaller electron tunneling in S3 results in a richer spectrum, exhibiting more maxima of radiation within the same frequency range, which is consistent with \Fig{fig:refl}(b) showing a denser polariton spectrum. At the same time, the maxima of THz emission from S2 are wider and more robust to changes of  the electric field $F$. In fact, for S3 the bend in the maximum at approximately $F=17.5$\,kV/cm is stronger than for S2, which is the result of the above mentioned sharper tunnel-induced anticrossing of both exciton and polariton lines, although there is no significant difference in the maximum intensity between the two systems. We also note that the THz emission intensity is always maximized when the two laser frequencies are in resonance with two different polariton states, as can be seen comparing Figs.\,\ref{fig:int}(a) and (d) with Figs.\,\ref{fig:refl}(a) and (b), respectively.

\begin{figure}
%    \centering
\vskip0mm
\hskip-15mm
\includegraphics[width=0.45\textwidth,angle=-90]{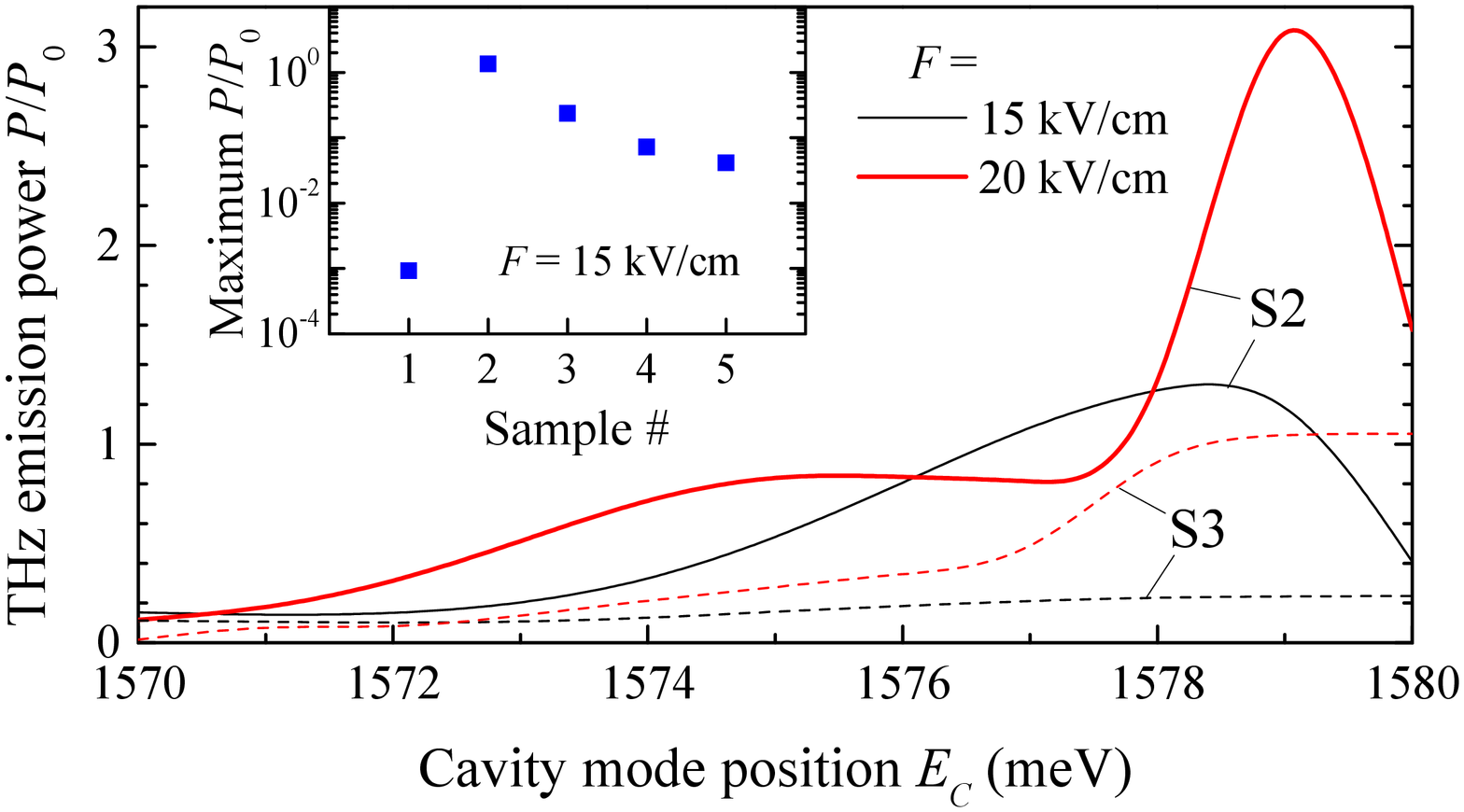}
\vskip-38mm
    \caption{Maximum  intensity of the THz emission from the DQW structures S2 (solid lines) and S3 (dashed lines) as functions of the cavity mode position, for $F=15.1$ (black) and 20.1\,kV/cm (red), and the lasers resonant to the two lowest dipolariton states. Inset: optimized over $\Delta$ and $E_C$ maximum intensity for $F=15.1$\,kV/cm, for samples S1--S5.}
    \label{fig:int_c}
\end{figure}

We also study in \Fig{fig:int_c} the dependence of the maximum intensity in both systems on the cavity mode position $E_C$, when the excitation laser frequencies $\omega_1$ and $\omega_2$ are matching the lowest two polariton lines, so that for each $F$ and $E_C$, the laser detuning $\hbar \Delta$ is given by the polariton splitting energy. For S2, the intensity peaks at $E_C\approx1579$\,meV for $F=15.1$\,kV/cm, demonstrating the optimal balance between the polariton splitting and the dipole moment, both contributing to \Eq{eq:singleXpower}. The trade-off between these two main quantities characterizing dipolaritons determines the optimal system for THz radiation. In fact, decreasing the barrier width in the DQW structure increases the polariton Rabi splitting, due to a larger tunnel coupling between the direct and indirect exciton, as it becomes clear comparing  Figs.\,\ref{fig:refl}(a) and (b). The tunnel coupling also potentially increases the overlap of the two polarizations $Y_1$ and $Y_2$ in \Eq{eq:dipolePol}. On the other hand, decreasing the barrier width, the dipole moment given by \Eq{eq:dipolePol} reduces, as the electron-hole separation is getting smaller. Vice versa, one could expect a larger dipole moment $d_{12}$ for wider DQW strictures, having a larger barrier separation, but this decreases the direct exciton component in the dipolariton, hence reducing the coupling to the cavity mode and, as a result, the Rabi splitting, thus leading to a reduced optimal detuning $\Delta$ (which is matching the polariton splitting) and a reduced power of THz radiation, in accordance with \Eq{eq:singleXpower}. This conclusion is illustrated by the inset in \Fig{fig:int_c} showing the maximum intensity of THz radiation from 5 different systems, all calculated at $F=15.1$\,kV/cm and the optimal cavity mode position ($E_C = 1571$\,meV for S1, $1579$\,meV for S2, $1580$\,meV for S3, $1575$\,meV for S4, and $1564$\,meV for S5). The inset also demonstrates that among the set of five systems discussed, S2 is the optimal one.

We see from \Fig{fig:int}(b) that the maximum power of THz emission in the investigated frequency range is $P\approx4P_0=3.2$\,mW, using the estimate for $P_0$ provided above. It is also of interest to estimate the
optical-to-THz conversion efficiency $\eta$ which is the coefficient of transformation of the energy of the excited polaritons within the MC into the THz radiation. The conversion efficiency can be evaluated as $\eta=P/P_E$, where $P_E$ is the polariton optical emission power due to a finite polariton lifetime $\tau$. The latter is determined primarily by the Q-factor of the cavity mode: $\tau=Q/\omega\approx3ps$, for the modeled systems. The emission power is then given by $P_E=\hbar\omega {\cal N}/\tau$, estimated to $P_E\approx 3.4$\,W for the number of polaritons ${\cal N}\approx4\times 10^7$ used above for the mesa of $0.1\times0.1$\,mm$^2$. We therefore find the THz efficiency $\eta\approx0.1\%$.  We note that this value of $\eta$ can be increased further by choosing a MC with a higher Q-factor, as well as by increasing the lateral size of the excited mesa.

Acknowledgements -- This work was supported by RFBR (Grant No. 16-29-03333) and Cardiff Undergraduate Research Opportunities Programme (CUROP).

\bibliography{letter}
\end{document}